\documentclass{article}
\title{The continuum problem and reality}
\author{O. Yaremchuk}
\date{\today}
\begin{document}
\maketitle
\begin{abstract}

Set-theoretical, physical, and intuitive notions of
continuum are compared. It is shown that the independence of
the continuum hypothesis determines status and properties of
the set of intermediate cardinality. The intermediate
set is a hierarchy of non-equivalent infinite sets.
Its description consist of autonomous theories.
In particular, quantum mechanics, classical mechanics,
and geometrical optics should be regarded as components of
the complete description of the intermediate set.
The intermediate sets may be produced as a result
of a fission of a continuous interval. The simplest
schemes of the fission are considered. Some analogy
with the Standard Model is pointed out.

\end{abstract}

\section{The continuum problem}

The primary inherent property of any finite set is the number
of the set members, i.e., its cardinality.
G. Cantor denoted cardinality by double bar above a symbol
of the set in order to indicate the double act of abstraction
from nature and order of the set members. Indeed, there is
no more to abstract from. It is the most fundamental
property determined by the most basic relation: equivalence.
The belief that this abstraction is possible, i.e., that
cardinality is independent of nature and arrangement of
a set members, is based on observation on finite sets.

In axiomatic set theory, infinite sets appear due to axiom
of infinity and power set axiom, which do not indicate any
difference between finite and infinite sets.
As a result, the supposed  independence was directly extended to
infinite sets but was not presented as an axiom: it seemed
too trivial.

Unlike a finite set, any infinite set is equivalent to its
proper subset. This means existence of a range of equivalence
inside any infinite set which is factually an internal symmetry
of the set. 

Note that the most symmetrical arrangement is the most probable
one because such an arrangement has the greatest thermodynamical
probability: the number of equivalent (symmetrical) states.
Thus free members of any infinite set should form the most symmetrical
arrangement determined by its cardinality.
In other words, any
infinite set cannot be arranged asymmetrically: it possess a
unique symmetry inherently (``by birth'') and this symmetry
reduces to the corresponding equivalence.

The tendency to symmetrization may be observed in large finite sets.
The concept of probability tells us that uncontrolled behavior of
a finite number of mutually independent objects of the same nature
becomes controlled by a symmetry when the number approach infinity.
The independent objects become more and more mutually dependent and
tend to form a symmetrical arrangement due to increase of cardinality
only. For example, if we throw points, independently, on
an continuous interval, configurations of a small number of
the points can be different and complicated but a large number
of independent throws gives more smooth configurations and we get
absolutely homogeneous outputs
when number of throwings $\to\infty$.\footnote{To some extent,
this example may serve as a justification for the point-set
approach to continuum in general.}

Some statistical laws, e.g. the law of large numbers,
and the concept of probability itself are based on
properties which should be regarded as pure set-theoretic
and related to the threshold between finite and infinite
cardinalities.

G. Cantor found only two infinite cardinal numbers: cardinality
of the countable set and cardinality of the set of all real numbers.
Since he did not found any set having more members than the
countable set and less than the set of all real numbers, he
supposed that such a set did not exist (the continuum hypothesis, CH).
Formalization of set theory has allowed to prove the independence
of CH \cite{Cohen}, i.e., undecidability of the continuum problem.
However, there is no interpretation of the independence.
Informally the continuum problem is not solved.

In order to clarify the status of the intermediate set,
the following two factors should be taken into account.
First, by definition, the set of intermediate
cardinality $M$ should be a subset of continuum $R$
(continuum should contain a subset equivalent to the intermediate set).
Second, separation of the intermediate set from continuum is a proof of
existence of the set and, therefore, is forbidden by the independence of the
continuum hypothesis. If we compare this points, we get that the
independence of CH should be understood as impossibility, in principle, to
separate the subset of intermediate cardinality from continuum.  In other
words, the independence of CH means that for any real number $x\in R$
the statement $x\in M\subset R$ is undecidable. We, in principle,
do not have a rule for separation of any subset with intermediate
number of members. Any separation rule for such a subset expressible in
Zermelo-Fraenkel set theory, ZF, is a proof of existence of the intermediate set
and, therefore, contradicts the independence of the continuum hypothesis, i.e.,
any rule (property) we can formulate implies separation of either continuous or
countable subset of $R$.

The inseparability may be explained by incompatibility between
symmetries of arrangements of the intermediate set and continuum.
We cannot localize the members of the intermediate
subset till they are in the arrangement of continuum. Relative
positions of the members of the intermediate set are different
from relative positions of the real numbers.

There is no ``labeling rule'' that allows separation
of the set of intermediate cardinality inside continuum. The
only way out is to separate it in the full sense: we should delete
all excess points outside the continuous set
(standard set theory does not forbid to
delete points one by one or by some portions)
and let the remaining points take the appropriate
arrangement.

To simplify the task, consider a continuous interval as
initial continuum instead of the set of all real numbers.
We cannot get  an intermediate interval out
of the continuous interval by decrease of the interval
length: all intervals with regular length are equipotent.
The three infinite numbers of points inside the interval and outside it
(left and right) remain equal and constant.
If the number of points of the interval begins to decrease, this means that
the number of points inside the interval becomes different from
the both outside numbers, i.e., two infinite sets of points get outside
the interval (left and right regions are mutually disjoint).

In order to
produce an interval of intermediate cardinality we should remove some
infinite set of points out of the continuous interval.  This set cannot be
continuous or countable. Elimination of a countable set cannot change
cardinality of continuum \cite{settheory}, i.e., the eliminated set should
have intermediate cardinality as well. In other words, we can get the
intermediate set by fission of a continuous interval $\bar{\alpha}$ into
intermediate (equivalent or non-equivalent) subintervals.

Introducing
operation of fission of continuum $\stackrel{fis}{\longrightarrow}$,
we can write the simplest scheme of the fission as follows:
\begin{equation}
\bar{\alpha}\stackrel{fis}{\longrightarrow}\bar{\gamma}_{\leftarrow}+\bar{\beta}+\bar{\gamma}_{\rightarrow}, \label{fission}
\end{equation}
where $|\bar{\gamma}|$ is a quantum of infinite cardinality
(cardinality of the minimum set $\bar{\gamma}$ extraction of
which is sufficient to change cardinality of continuum).
Thus, in ZF terms, we can get the intermediate set through
the stage of continuum, by the backward step:
$N\stackrel{2^N}{\longrightarrow}R\supset\bar{\alpha}\stackrel{fis}{\longrightarrow} I$.

In order to get smaller intermediate cardinality, we need to press the next
pair of quanta out of the interval and make the more rarefied structure.
Thus infinite cardinalities are quantized: they decrease by steps with
``emission'' of some infinite quanta of points over the end points of the
interval.

Till the processed interval has cardinality of continuum (the number
of deleted points is no more than countable), its points can
be put into one-to-one correspondence with points of some whole interval,
which is continuous in ordinary sense. This correspondence may be regarded
as an rearrangement of the points of the rarefied interval into the complete
interval. Due to this fundamental rearrangement, we do not need to use
the concept of measure and touch upon related problems
(covering, countable additivity, etc.).
This is important, since most of the habitual mathematical
notions may be ``infected'' with the continuum hypothesis,
i.e., they can implicitly contain properties equivalent to CH.

The output intermediate intervals also cannot be considered as
discontinuous, fragmented intervals. The intermediate interval cannot
contain discontinuity points (``holes'' from removed points) as it is
always implied in ordinary calculus. The remaining points should
rearrange into some new whole (gapless) structure.
Thus the intermediate set is some other type of
continuum which may be called incomplete or structured continuum.

Since ordinary discontinuity criterion does not work, intermediate
cardinality of the incomplete continuous interval should
reveal itself only in the behavior of its length.
All continuous intervals have regular length as manifestation
of the equivalence to the set of the real numbers, while length of
a rarefied intermediate interval must show some irregularity.
The ``sieve'' interval cannot shrink into a regular interval,
its length should be affected (damaged or destroyed).
In more general terms, such an interval is not self-congruent.

At some step of rarefication of the interval, its length as
a manifestation of continuity should vanish, i.e.,
eventually, the interval should turn to point identically.
The non-self-congruent intervals have quite expected
transitional cardinality from regular to zero length.

It is relevant here to recall long-term controversy between
discrete and continuous notions of space. These notions are not pure
mathematical abstract concepts. The problem is the identification
of the fundamental physical object, the basic element of reality.
Obviously, the set of intermediate cardinality can be neither
continuous nor discrete. Thus the continuum problem is
related to the concrete  fundamental object.

\section{Quantum mechanics}

Coordinate of a particle at any instant is length of the
interval between the particle and a reference point.
In classical mechanics, we may take the coordinate interval
as a unit of measurement.
This means that the coordinate of the point
is {\it a priori} equal to $1$ exactly. We do not need to
measure the length etalon, we only need to keep it congruent
to itself.

In quantum mechanics, the choice of the unit does not cancel
quantum indeterminacy and does not allow to avoid measurement
and get {\it a priori} coordinate. On the other hand, quantum
particle is point-like and this point is in space permanently,
i.e., the coordinate interval of a quantum particle always
exists but it has no definite length before measurement.
In other words,
coordinate interval of a quantum particle before measurement is
not self-congruent and cannot be used as a unit of length
(it cannot be used even to measure itself).
Measurement of a quantum particle coordinate is an identification
of a non-self-congruent interval and some self-congruent interval
(existence of exactly self-congruent intervals is a fact too).

This is not an interpretation but
the basic and elementary  property of the microscopic
geometry which may be reduced only to more elementary
set-theoretic factors. It is unnecessary and baseless to state that
the coordinate interval of a quantum particle does not exist
before measurement.

Self-congruence of intervals is included in the first congruence
axiom of the Hilbert's axiomatics of geometry as an addition.
Some authors omit this addition as too self-evident. The Euclid's
axiomatics has no axiom of congruence. Thus geometry with
non-self-congruent intervals may be called non-Hilbertian.
Note that the axioms of congruence imply the continuum hypothesis.

Congruence could be a key word for understanding quantum
mechanics (Hilbert's "Grundlagen der Geometrie" was published in 1899)
but, at the beginning of the 20th century, focus was on the fifth Euclid's
postulate.

\section{The continuum problem and quantum\\mechanics without
interpretation}

Consider maps of the intermediate set $I$ to
the sets of real numbers $R$ and natural numbers~$N$:
\begin{equation}
N\gets I\to R.
\end{equation}

Let the map $I\to N$ decompose $I$ into
the countable set of mutually disjoint
infinite subsets: $I\to \{I_n\}$ ($n\in N$).
Let $I_n$ be called a unit set. All members of $I_n$
have the same countable coordinate $n$.

Consider the map $I\to R$.
By definition, continuum $R$ contains a subset $M$ equivalent to~$I$,
i.e., there exists a bijection
\begin{equation}
f:I\to M\subset R.
\end{equation}
This bijection reduces to separation of the intermediate subset $M$
from continuum. Since one cannot distinguish members of $M$
from the remaining real numbers (there is no formulation of a rule ot a
property that ``marks'' the intermediate number of members), each
member of the set of intermediate cardinality equally corresponds
to all real numbers. It is important that we consider mapping of
the isolated intermediate set $I$ in its special arrangement.
The bijection, as an equivalence relation, instead of expected separation,
establishes equivalence between any member of $I$ and all real numbers.
If we could localize of the intermediate set points,
coordinates of the points were self-congruent intervals.

Thus only random (arbitrarily chosen) real number can be
assigned to an arbitrary point $s$ of the intermediate set.

In the case of inexact measurement of a self-congruent interval,
possible results are not equivalent. There is a deviation from
exact length evaluating the results and making them non-equivalent.
For a non-self-congruent interval,  appropriate ''measuring procedure''
should ensure equiprobability of equivalent real numbers and
unbiased choice of a unique number by a suitable random process.

In quantum mechanics we also have rather some kind
mapping than measurement of length in the classical
sense: ordinary measurement operations are impossible in
principle and the output real number cannot be understood
classically.

After the choice has performed,
a concrete point gets a random real number as its
coordinate in continuum.
Thus we get probability $P(r)dr$ of finding the
point $s\in I$ about $r$.

Thus the point of the intermediate set has two coordinates:
definite natural number and random (arbitrarily chosen) real number:
\begin{equation}\label{s}
s:(n,r_{random}).
\end{equation}

Only the natural number coordinate gives reliable
information about relative positions of the points
of the set and, consequently, about size of an interval.
But the points of a unit set are indistinguishable.
Without loss of generality we shall use a fixed
countable mapping.

For two real numbers $a$ and $b$ the probability
$P_{a\cup b}dr$ of finding $s$ in the union of the
neighborhoods $(dr)_a\cup (dr)_b$
\begin{equation}
P_{a\cup b}\,dr\ne [P(a)+P(b)]\,dr
\end{equation}
because $s$ corresponds to all real numbers and, therefore, to both
neighborhoods  at the same time (the events are not independent).
It is most natural, in this case, to compute the non-additive probability
from some additive object by a simple rule. Since the point corresponds
to all real numbers simultaneously, we may associate with the point  a function
$\psi(r)$ defined on the same domain $R$ such that $P(r)={\cal P}[\psi(r)]$
and $\psi_{a\cup b}=\psi(a)+\psi(b)$.
It is quite clear that the dependence ${\cal P}[\psi(r)]$ should be non-linear.
Indeed,
\begin{equation}
P_{a\cup b}={\cal P}(\psi_{a\cup b})={\cal P}[\psi(a)+\psi(b)]\ne
{\cal P}[\psi(a)]+{\cal P}[\psi(b)].
\end{equation}
We may choose the dependence arbitrarily but the simplest
non-linear dependence is the square dependence:
\begin{equation}
{\cal P}[\psi(r)]=|\psi(r)|^2.
\end{equation} \label{born}

We shall not discuss uniqueness of the chosen options.
Our aim is to show that quantum mechanics may be reproduced as
a natural descriptions of the set of intermediate cardinality.
There is no need to insist that it is a unique or the best way to
describe the set.

The function $\psi$, necessarily, depends on $n$: $\psi(r)\to\psi(n,r)$.
Since $n$ is accurate up to a constant (shift)
and the function $\psi$ is defined up to the
factor $e^{i\mbox{const}}$, we have
\begin{equation}
\psi (n+\mbox{const},r)=e^{i\mbox{const}}\psi (n,r).\label{gauge}
\end{equation}
Hence, the function $\psi$ is of the following form:
\begin{equation}
\psi (r,n)=A(r)e^{2\pi in}\label{debr}.
\end{equation}
Our choice of the non-linear dependence allows to ensure the invariance
under shift in $N$ (and consequently in $I$) and we do not need to revert
and find more suitable dependence. For our purpose, it is not
necessary to prove that this is a unique dependence satisfying the
requirements of the non-linearity and the invariance and to replace
Born postulate by the uniqueness theorem. Note, however, that the pure
postulation is replaced by the choice.

Thus the point of the intermediate set corresponds to
the function Eq.(\ref{debr}) in continuum. We can specify
the point by the function $\psi(n,r)$ before the mapping
and by the random real number and the natural
number when the mapping has performed. In other words,
the function $\psi(n,r)$ may be regarded as the image
of $s$ in $R$ between mappings.

Consider probability $P(b,a)$ of finding the point $s$ at $b$
after finding it at $a$.
Let us use a continuous parameter $t$ for correlation
between continuous and countable coordinates of the point
$s$ (simultaneity) and in order to distinguish between
different mappings (events ordering):
\begin{equation}
r(t_a),n(t_a)\to\psi(t)\to r(t_b),n(t_b),
\end{equation}\label{0-t}
where $t_a<t<t_b$ and $\psi(t)=\psi[n(t),r(t)]$.
For simplicity, we shall identify the parameter with
time without further discussion. Note that we cannot use the
direct dependence $n=n(r)$: since $r=r(n)$ is a random number,
the inverse function is meaningless.

Assume that for each $t\in (t_a,t_b)$ there exists the image of
the point in continuum $R$.

Partition interval $(t_a,t_b)$ into $k$ equal parts
$\varepsilon$:
\begin{eqnarray}
k\varepsilon =t_b-t_a,\nonumber\\
\varepsilon =t_i-t_{i-1},\nonumber\\
t_a=t_0,\,t_b=t_k,\\
a=r(t_a)=r_0,\, b=r(t_k)=r_k.\nonumber
\end{eqnarray}\label{partition}
The conditional probability of of finding the point $s$ at
$r(t_i)$ after $r(t_{i-1})$ is given by
\begin{equation}\label{cond}
P(r_{i-1},r_i)=\frac{P(r_i)}{P(r_{i-1})},
\end{equation}
i.e.,
\begin{equation}
P(r_{i-1},r_i)=\left|\frac{A_i}{A_{i-1}}e^{2\pi i\Delta n_i}\right|^2,
\end{equation}
where $\Delta n_i=|n(t_i)-n(t_{i-1})|$.

The probability of the sequence of the transitions
\begin{equation}
r_0,\ldots ,r_i,\ldots r_k
\end{equation}\label{sequence}
is given by
\begin{equation}
P(r_0,\ldots ,r_i,\ldots r_k)=\prod_{i=1}^k P(r_{i-1},r_i)=
\left|\frac{A_k}{A_0}\exp 2\pi i\sum_{i=1}^k\Delta n_i\right|^2.
\end{equation}
Then we get probability of the corresponding continuous sequence
of the transitions $r(t)$:
\begin{equation}\label{pathprob}
P[r(t)]=\lim_{\varepsilon\to 0}P(r_0,\ldots ,r_i,\ldots r_k)=\left|\frac{A_k}{A_0}e^{2\pi im}\right|^2,
\end{equation}
where
\begin{equation}
m=\lim_{\varepsilon\to 0}\sum_{i=1}^k\Delta n_i.
\end{equation}

Since at any time $t_a<t<t_b$ the point $s$ corresponds to all points
of $R$, it also corresponds to all continuous random sequences of
mappings $r(t)$ simultaneously, i.e., probability $P[r(t)]$ of finding
the point at any time ${t_a\leq t\leq t_b}$ on $r(t)$ is non-additive too.
Therefore, we introduce an additive functional $\phi[r(t)]$.
In the same way as above, we get
\begin{equation}
P[r(t)]=|\phi[r(t)]|^2.
\end{equation}\label{}
Taking into account Eq.(\ref{pathprob}), we can put
\begin{equation}\label{phi}
\phi[r(t)]=\frac{A_N}{A_0}\,e^{2\pi im}=\mbox{const}\,e^{2\pi im}.
\end{equation}
Thus we have
\begin{equation}\label{pathsum}
P(b,a) = |\!\!\sum_{all\,r(t)}\!\!\mbox{const}\,e^{2\pi im}|^2,
\end{equation}
i.e., the probability $P(a,b)$ of finding the point $s$ at $b$
after finding it at $a$ satisfies the conditions of Feynman's approach
(section 2-2 of \cite{Feynman}) for $S/\hbar=2\pi m$.
Therefore,
\begin{equation}
P(b,a)=|K(b,a)|^2,
\end{equation}
where $K(a,b)$ is the path integral (2-25) of \cite{Feynman}:
\begin{equation}\label{pathint}
K(b,a)=\int_{a}^{b}\!e^{2\pi im}D r(t).
\end{equation}
Since Feynman does not essentially use in Chap.2 that $S/\hbar$ is just
action, the identification of $2\pi m$ and $S/\hbar$ may be postponed.

In section 2-3 of \cite{Feynman} Feynman explains how
the principle of least action follows from the dependence
\begin{equation}\label{sum}
P(b,a)= |\!\sum_{all\,r(t)}\!\!\mbox{const}\,e^{(i/\hbar)S[r(t)]}|^2.
\end{equation}
This explanation may be called ``Feynman's correspondence principle''.
We can apply the same reasoning to Eq.(\ref{pathsum}) and,
for very large $m$, get ``the principle of least $m$''.
This also means that for large $m$ the point $s$ has a definite
(self-congruent) path and, consequently, a definite continuous coordinate.
In other words, the interval of the intermediate set with the large
countable length $m$ is sufficiently close to continuum (let the
interval be called macroscopic), i.e., cardinality of the intermediate
set depends on its size.

For sufficiently large $m$,
\begin{equation}\label{min}
m=\int_a^b\!\! dm(t)=\int_a^b\!\! \frac{dm(t)}{dt}\,dt=\min.
\end{equation}
Note that $m(t)$ is a step function and its time derivative is almost
everywhere exact zero. But for sufficiently large increment
$dm(t)$ the time derivative  ${\frac{dm}{dt}=\dot{m}(t)}$
makes sense as non-zero value.

The function $m(t)$ may be regarded as some function of
$r(t)$: $m(t)=\eta[r(t)]$. It is important that $r(t)$ is not
random in the case of large $m$. Therefore,
\begin{equation}\label{f}
\int_a^b\!\!dm(t)=\int_a^b\!\frac{d\eta}{dr}\,\dot{r}\,dt=\min,
\end{equation}
where $\frac{d\eta}{dr}\,\dot{r}$ is some function of $r$, $\dot{r}$,
and $t$. This is a formulation of the principle of least action
(note absence of higher time derivatives than $\dot{r}$), i.e.,
large $m$ can be identified with action. Recall that this identification is
valid only for sufficiently large $dm=\dot{m}dt$, i.e., for sufficiently fast
points in sense of time rate of change of the countable coordinate.

Feynman's correspondence principle indicates a qualitative leap
which is inverse to quantization: action is not the length of the countable
path but some new function.
We get a new characteristics of the point and a new
law of its motion. In modern terms, this is the phenomenon
of decoherence explained without referring to environment.
It is possible to do with internal reasons (e.g. some ``microscopic
explosion'', etc.).
In contrast to usual decoherence, it is able to explain appearance
of classical universe from its early quantum stages.

Since the value of action depends on units of measurement, we need
a parameter $h$ depending on units only such that
\begin{equation}
hm=\int_a^b\!\! L(r,\dot{r},t)\,dt=S,
\end{equation}
were $ L(r,\dot{r},t)=\frac{d\eta}{dr}\,\dot{r}$ is
the Lagrange function of the point.

Finally, we may substitute $S/\hbar$ for $2\pi m$ in Eq.(\ref{pathint})
and regard our consideration as an extension of Feynman's formulation
of quantum mechanics which simplifies the original Feynman's
approach because there is no need in classical paths and existence of
action from the very beginning.

Note that if time rate of change of cardinality (i.e., of the countable
coordinate) is not sufficiently high, action vanishes: $\dot{m}(t)$ and,
consequently, ${dm=\dot{m}(t)dt}$ is exact zero. This may be understood
as vanishing of the mass of the point. Formally, mass is a consequence of
the principle of least action: it appears in the Lagrangian of a free
particle as its specific property \cite{mech}. Thus mass is somewhat
analogous to air drag which is substantial only for sufficiently fast
bodies.

Consider the special case of constant time rate of change $\nu$ of
the countable coordinate $n$ (since $n$ is a natural number, we can always
divide $(t_a,t_b)$ into intervals of constant $\nu$). We have $m=\nu (t_b-t_a)$. Then
``the principle of least $m$'' reduces to ``the principle of least $t_b-t_a$''.
If $\nu$ is not sufficiently large (massless point), this is the
simplest form of Fermat's least time principle for light. The more
general form of Fermat's principle follows from Eq.(\ref{min}): since
\begin{equation}
\int_{t_a}^{t_b}\!\!dn(t)=\nu\!\int_{t_a}^{t_b}\!\!dt=\min,
\end{equation}
we obviously get
\begin{equation}\label{fermat}
\int_{t_a}^{t_b}\!\!\frac{dr}{v(t)}=\min,
\end{equation}
where $v(t)=dr/dt$.
In the case of non-zero action (mass point), the principle
of least action and Fermat's principle ``work''
simultaneously. It is clear that any additional factor (mass, in
this case) can only increase the pure least time. As a result,
$t_b-t_a$ for a massless point bounds below $t_b-t_a$
for any other point and, therefore, $(b-a)/(t_b-t_a)$
for massless point bounds above average speed between
the same points $a$ and $b$ for continuous image of any
point of the intermediate set.
Thus, paradoxically, the countably slowest points
have the fastest continuous image. Recall that
any spacetime interval along a light beam is exact zero
(lightlike or null interval). In this sense, a photon,
in Minkowski spacetime, is at absolute rest. Recall also that
the spacetime interval is directly related to the relativistic
action.

On the contrary, a point with mass should permanently have sufficiently
high time rate of change of its path cardinality. As a consequence, the
continuous image of the point cannot be fixed as well. This quite conforms
with the uncertainty principle which is rather ``non-stop principle'.
It is worthwhile to stress drastic distinction between classical and
quantum parts of reality: perpetual motion is the ordinary motion in
the quantum world. There is no reference frame in which a quantum
particle is at rest. This is true not only for a photon (which underlies
the special relativity) but for mass particles too.

We should also expect existence of points with transitional time rate of
change of cardinality for which mass is not stable. In this case, mass
should vary between exact zero and small, discrete non-zero values.
The spectrum of masses should be discrete, since it depends on the
countable speed and its threshold values.

Note that recently discovered neutrino oscillations imply necessarily
oscillations of their masses.
Automatically, we get explanation of the neutrino helicity:
frame of reference, which is faster than such a point, obviously,
cannot be found (only massless points are faster) and
linear momentum of the point cannot be reversed.

\section{The intervals}

Thus we get three kinds of the intervals:

\begin{enumerate}

\item macroscopic continuous intervals that have regular length as
a manifestation of their equivalence to the set of the real numbers;

\item submicroscopic intervals with unstable lengths;

\item microscopic intervals without length which are,
actually, composite points.

\end{enumerate}

A submicroscopic interval is a non-self-congruent {\it extended}
interval (uncompleted continuum). Its direction is coupled to the direction of the
parent continuum.

The intervals are regular, irregular and zero-length coordinates of
true and composite points.

If the natural number constant (shift) in Eq.(\ref{gauge}) is
bound within a composite point $\bar{s_c}$, it becomes
dependent on the point and its structure: $\mbox{const}\to n'(r_c)$.
Then
\begin{equation}
\psi_c (n_c,r_c)=e^{in'(r_c)}\psi_c (n_c,r_c),
\end{equation}\label{phase}
where $n_c$ is the natural number coordinate of any point of
the interval.
Clearly, this is the origin of the gauge principle.
This means that sufficiently close but microscopically
distinguishable (non-equivalent) points become exactly equivalent
macroscopically, i.e., it is actual conflict of equivalences.
On the other hand, gauge symmetry may be regarded as a
phenomenological confirmation of existence of the point-like
intervals.

Due to the properties of non-equivalent infinite sets,
one-dimensional intermediate axis splits into, at least, three
non-equivalent subaxes,  i.e., ``immiscible'' substructures with
different internal symmetries. The complete description is
three-dimensional.

In this case, dimensionality is a classification of cardinalities.
The classification with respect to length is the roughest (macroscopic)
estimation of cardinality (yes, no, unstable). Length is an
indication of the degree of saturation of cardinality: saturated (continuum),
unsaturated, close to saturation. Thus we get the following spectrum:
\begin{equation}
|R|>|I_{\sim}|>|I_{0}|>|N|,
\end{equation}\label{spectrum}
where $I_{\sim}$ and $I_{0}$ are the sets of the submicroscopic
and proper microscopic intervals respectively.

Since the microscopic intervals are essentially non-equivalent, they
form a quantity of different autonomous objects. Therefore, description
of the structure and transmutation of the intervals needs additional
dimensions down to the single unit set. But these dimensions should
manifest themselves rather as qualitative properties (charges) of the
composite points.

A moving composite point, in one-dimensional case, forms (formally)
a countable two-dimensional ``sheet''. For a sufficiently fast point,
it gives a continuous variable (action) which may be interpreted
as {\it dequantized} area of the countable surface.

In the fission Eq.(\ref{fission}), we have the ``point-antipoint'' pair
$(\bar{\gamma}_{\leftarrow}, \bar{\gamma}_{\rightarrow})\in I_{0}$
separated by the unstable continuous interval $\bar{\beta}\in I_{\sim}$
Here we regard a point moving in the direction opposite to
the positive direction of the countable axis as an ``antipoint''.
Recall that we use a fixed countable mapping.

In order to get three point-like intervals as a result of the fission,
we need one more set of proper microscopic intervals in the spectrum of
infinite cardinalities:
\begin{equation}
|R|>|I_{\sim}|>|I^2_{0}|>|I^1_{0}|>|N|.
\end{equation}\label{spectrum1}
If all three intervals are point-like, we get a more
complex point consisting of three inseparable composite
points which are members of $I^2_{0}$. In this case, there are the
following possibilities. For countably motionless decomposing
interval, we also have the ``point-antipoint'' pair but
separated by one more point-like interval:
\begin{equation}
(\gets,-,\to).
\end{equation}\label{meson}
For sufficiently fast interval:
\begin{equation}
\overrightarrow{(\gets,\to,\to)},\overrightarrow{(\gets,\gets,\to)}
\end{equation}\label{right}
and
\begin{equation}
\overleftarrow{(\gets,\to,\to)},\overleftarrow{(\gets,\gets,\to)},\label{left}
\end{equation}
where arrows in the brackets mean zero-length motions of component
points (analogue of asymptotically free motion) which are possible
in combination with motion of each three-component point as a whole.
Obviously, free zero-length motion cannot destroy the complex point.
Time rate of change of infinite cardinality has the same discrete
spectrum, i.e., it can change by multiples of the infinite ``portions''.

In this simplest one-dimensional case, one nevertheless can see
obvious analogy with meson and nucleon structures of the Standard
Model including some features that have provoked spin crisis:
the intergral countable motion (``spin'') of the ``nucleons'' is not
a combination of the individual motions of ``quarks'' and some
individual motions are opposite to the integral one.
Then $(\to)$ should be regarded as an up quark and $(\gets)$ as a
down quark (in the right moving objects).
Thus, preliminarily, $\overrightarrow{(\gets,\gets,\to)}$ may
be identified with neutron and $\overrightarrow{(\gets,\to,\to)}$
with proton:
\begin{equation}
\overrightarrow{(\gets,\gets,\to)}=ddu,\,
\overrightarrow{(\gets,\to,\to)}=duu.
\end{equation}
As a result of combination of speeds,
all component points of ``nucleons'' move right,
while the ``antinucleons'' in Eq.(\ref{left}) consist of ``antiquarks''
(left resultant speed). It seems to be meaningless to compare masses
of the components.

Note that a composite point, in contrast to a true point,
cannot have arbitrary countable speed.
Cardinality of the composite point determines its level of
propagation (range of equivalence) and, consequently,
minimum non-zero time rate of change of its path cardinality.

The correct description of composite objects needs  taking into
account many factors: dimensionality, complete spectrum of cardinalities,
all ways of fission and recombinations of the fragments, all kinds and
combinations of countable motions of the objects and their components,
etc.

The straightforward taking into account dimensionality leads to
trebling of objects.
We get three kinds of simple composite points, components of
complex points and three kinds of extended but unstable submicroscopic
intervals (fields).

It should be noted that cardinality does not change
not only in the case of a countably motionless point but also in
the case of a point with sufficiently high saturated countable
speed. This threshold speed obviously decreases with increase of
the countable length (cardinality) of a composite point.

Microscopic phenomena should be considered rather in the
countable subspace and in its dequantized ``version''
(spacetime) than in the macroscopic continuous space.

\section{Conflicts}

Feynman's correspondence principle does not reduce classical
mechanics to more precise quantum mechanics but separates
their regions of validity and explains the interrelation and
the difference between their dynamical laws.
Thus classical and quantum mechanic cannot be made
equivalent by any correspondence principle, i.e.,
the macroscopic world is not an extension of the
microscopic one.\footnote{If quantum mechanics was
universaly valid, all distances were non-self-congruent.
However, existence of exactly self-congruent distances is a fact.}
Hence, under some conditions,
classical and quantum mechanics can conflict.
Since classical and quantum systems are radically different in
power (cardinality) their direct contact results in suppression
of some features of the weaker system (phenomenon of
decoherence).

Theoretically, conflict is not a superposition of different
factors acting on the same object but a contradiction between
correct descriptions (``programs'') determining behavior of
the object.

The measurement problem and the problem of the wave function
collapse cannot be solved in the expected formally ideal
sense, since the measurement and the collapse are basically conflicts.
The general feeling of dissatisfaction (accompanying quantum mechanics
and its interpretations from the very beginning) turns out to be
unavoidable and even relevant when dealing with discordant
microscopic world.
It also seems to be impossible to reconcile completely point-set
and wholeness formal approaches to continuum.

\section{Life}

The more or less objective concept of conflict can be found only
in cybernetics, where it is connected, in particular, with common use of
the same resources by different systems or programs that may result in
incompatible properties or requirements for some objects.
Recall that, according to N. Wiener , cybernetics
is ``control and communication in the animal and the machine.''
Indeed, the functioning of the living matter and man-made systems
is a field of reality where conflicts are most apparent as natural phenomena.
Therefore, it is reasonable to expect that the living matter is the region
where the parallel descriptions interact by means of some interface that
prevents decoherence and makes possible substantial action of very weak
system on much more powerful one. It is clear that the region should be
accompanied with conflicts which are  indications of involvement of
discordant laws.

Strictly speaking, in the uniform world which can be described by a
unique theory of everything, there is no formally explainable origin
of conflicts. Without fundamental contradictions, it is not possible to
overcome subjectivity of the concept of conflict, i.e., conflicts cannot
be regarded as natural phenomena.

At present, the living matter is regarded as a very rare, random formation
without any special reason for its existence. On the contrary, the region
of correlation of the parallel substructures is a fundamental spatial
phenomenon.

The regulation of conflicts is possible only if very weak influence of
extremely small system can affect much more numerous, powerful system
which is environment for the
microscopic one due to difference of sizes and position in the
hierarchy of structures. A stable trend to complication and extraordinary
increase in complexity should be expected as a sign of such an influence.
The regulation leads to the conception of control as an inherent
property of the superstructure. In this case, control is domination of the
weaker factor, instead of composition of factors inherent to the inanimate
matter.

Some special interface is necessary in order to transfer the weak
influence of the small substructure and to prevent destructive impact
of more powerful structure.
Such an interface is a key factor for possibility of the correlated
region.
It is clear that the interface should be a part of the more
powerful subsystem (recall macroscopic measuring apparatus).

In the living matter, the interface may be formed, for instance,
by spatial configurations of organic macromolecules which
are definitely macroscopic formations.
It is well known that proteins, the necessary basic
components of all known forms of life, have rigid
spatial structures directly connected to their biological properties
and functions. Some small changes of the protein spatial structure
have important biological functions. Violation of the spatial structure
leads to inability to functionate.

Note that, in the two-slit experiment,
which may be regarded as a hint at the interface,
we also have a rigid spatial configuration (the slit width,
the inter-slit distance, the distance between the slits and the screen),
the violation of which destroy the interference pattern.

Pressing-out infinite number of points by an inexpressible rule for it,
radically complicates the configuration of the proper microscopic
interval. Figuratively speaking,
microscopic structures are cut out of the whole ``piece''
and then assembled into constructions.
Paradoxically, complexity of a ``monolithic'' microscopic object is,
in principle, inaccessible for a macroscopic structure of any size
consisting of finite
number of parts in a regular arrangement (order) or in the state of
chaos (disorder) which simplifies the configurations as well.

It is possible to introduce the concept of infinite complexity of the
non-constructible (non-assembled) system formed in the negative manner,
uniquely connected with the internal symmetry of the corresponding
infinite set. An infinite set can have infinite complexity: recall
that infinite sets are structures.

The level of complexity and organization is determined by depth of
correlation of substructures rather than by ``horizontal'', extensional
complexity of the constructible macroscopic substructure. This conclusion
has an interesting biological confirmation.
Human genome is not as different from that of the most primitive organisms
as it is reasonable to expect. If we compare this fact with the
apparent fact that the human behavior is incomparably more complicated
than behavior of any animal (culture, technology, etc.), we get that the
difference in complexity of the behavior should be considered as an
indication of presence of a directly undetectable additional controlling
structure of high complexity in human species which is not included in
the description (genome).

Complexity of a growing organism increases without
natural selection. Genome is regarded as the only ``program''
for the personal ``evolution'' (the central dogma of molecular
biology). But the difference in complexity between
the human and, for instance, the worm's genome is not
sufficient to ensure the observed difference in complexity of
their behaviors under any method of evaluation of the complexities.

Directly undetectable subsystem can manifest itself only
in the behavior of the supersystem (its directly perceptible
macroscopic part). The behavior of a living object is, on the
one hand, unpredictable and, on the other hand, non-random.
This is the necessary sign of inaccessible complexity.

Note that the variety of organisms decreases with increase of their level
of complexity: there is a great number of simple organisms but the unique
most complex one: human being. This is one more confirmative fact: the
finite number of different infinite cardinalities decreases from very
large number of the unite sets of the smallest continuous interval (the
smallest self-congruent interval) down to the single unite set.
In case of extensive, combinatorial complexity, the situation would be
inverse: the variety of more complex forms should be greater.

Thus the general biological arguments for the hypothesis of life as a
correlated region are

\begin{quote}

conflicts;

discreteness of the spectrum of species (absence of intermediate forms);

pyramidal form of the spices hierarchy (mirror analog of
inverted pyramid of infinite cardinalities);

inexplicably large difference in complexity of behavior of
genetically close species.

\end{quote}

Without environment of the less complicated species the complex
species cannot exist. The simple creatures also adapted themselves to
use the complex ones after they had appeared. Thus pyramid of species
developed as a whole: ``unfit'', simpler species did not vanish but
mutated to form necessary complexity levels of biosphere (``upward''
correlation as a constructible result of the correlation deepening).

It is sufficiently evident that ``vertical'' biological evolution is
completed, although mechanism of mutation and selection still works.

The most remarkable conclusion is that any living object can contain
a subtle autonomous part which, in principle, can remain safe
after destruction or detachment of the macroscopic substructure (body).

\section{Some remarks}

Instead of abstract fundamental principles,
we get the fundamental object and its properties.
This removes the problem of interpretation which is,
in essence, authentication of the object under investigation.

The intermediate set gives convincing informal picture and
generates the same peculiar concepts that have been introduced into
quantum theories for consistency with experiments including the
concepts without classical analog.
The structure of the formal description of the set follows progressively
the structure of fundamental physics which
is directly related to the structure of physical reality.

One can find in the mathematical literature a lot of statements
which are declared to be independent of cardinalities of sets
under consideration. However, in the presented framework, it
becomes clear that the true independence cannot be easily
achieved. The most of basic geometric and algebraic notions
inherently depend on cardinality of complete continuum
stronger than it could be expected and, therefore, imply
cardinality induced hidden postulates which are, in essence,
equivalent to the continuum hypothesis.
For instance, presence of the microscopic 10-dimensional
objects in spacetime is regarded as identical to 10-dimensionality
of spacetime as a whole. In fact, presence of microscopic
multidimensional objects  inside four-dimensional spacetime is
possible without radical change of full-scale spacetime structure.
There is no need to introduce six macroscopic, continuous
extra dimensions and then to hide them by one more
radical change: compactification.
Only correct informal understanding of reality can prevent
a formal description from fatal logical errors.

\end{document}